\documentclass[sigconf, review=false, anonymous=false]{acmart}

\setcopyright{none}
\renewcommand\footnotetextcopyrightpermission[1]{}

\acmConference[Vision Paper]{}{November}{2022}

\begin{document}

\title{Rethinking Storage Management for Data Processing Pipelines in Cloud Data Centers}
\subtitle{Vision Paper}

\author{Ubaid Ullah Hafeez}\authornote{Work done while at Google.}
\affiliation{%
  \institution{Stony Brook University}
  \city{Stony Brook}
  \state{NY}
  \country{USA}
  \postcode{11794}}
\email{uhafeez@cs.stonybrook.edu}

\author{Martin Maas}
\affiliation{%
  \institution{Google Reseaerch}
  \city{Mountain View}
  \state{CA}
  \country{USA}}
\email{mmaas@google.com}

\author{Mustafa Uysal}
\affiliation{%
  \institution{Google}
  \city{Mountain View}
  \state{CA}
  \country{USA}}
\email{uysal@google.com}

\author{Richard McDougall}
\affiliation{%
  \institution{Google}
  \city{Mountain View}
  \state{CA}
  \country{USA}}
\email{richardmcd@google.com}

\begin{abstract}
Data processing frameworks such as Apache Beam and Apache Spark are used for a wide range of applications, from logs analysis to data preparation for DNN training. It is thus unsurprising that there has been a large amount of work on optimizing these frameworks, including their storage management. The shift to cloud computing requires optimization across all pipelines concurrently running across a cluster. In this paper, we look at one specific instance of this problem: placement of I/O-intensive temporary intermediate data on SSD and HDD. Efficient data placement is challenging since I/O density is usually unknown at the time data needs to be placed. Additionally, external factors such as load variability, job preemption, or job priorities can impact job completion times, which ultimately affect the I/O density of the temporary files in the workload. In this paper, we envision that machine learning can be used to solve this problem. We analyze production logs from Google's data centers for a range of data processing pipelines. Our analysis shows that I/O density may be predictable. This suggests that learning-based strategies, if crafted carefully, could extract predictive features for I/O density of temporary files involved in various transformations, which could be used to improve the efficiency of storage management in data processing pipelines.
\end{abstract}

\maketitle

\section{Introduction}
\label{s:intro}

Data processing frameworks such as Apache Beam~\cite{apache_beam} or Apache Spark~\cite{zaharia2010spark} are used for a wide range of applications, from log analysis to preparing data for training deep neural networks (DNNs). Collectively, these data processing pipelines consume large amounts of compute and storage resources. It is therefore unsurprising that these frameworks have seen a large amount of research optimizing diverse aspects of their execution, ranging from scheduling \cite{rao2012sailfish, yang2019apachenemo} to data placement \cite{stuedi2017crail,rasmussen2012themis,feng2016optimization}.

Most of this prior work focused on individual pipelines in isolation and aims to reduce their resource utilization (e.g., by running on a single machine or a small cluster). However, the shift to cloud computing means that a very common, if not the most common, execution model of these pipelines is to run in shared data centers \cite{isard2007dryad,yang2019apachenemo,rao2012sailfish,condie2010mapreduce}. The workload thus interacts with a range of shared distributed infrastructure that is often ignored in prior research.

In this paper, we focus on one such aspect – the storage layer. When running in data centers, data processing pipelines often read and write data to distributed file systems \cite{yang2019apachenemo, condie2010mapreduce, rao2012sailfish, stuedi2017crail} such as HDFS~\cite{shvachko2010hdfs}, GFS~\cite{sanjay2003googlegfs}, or other shared storage infrastructure. This gives rise to a new set of problems and trade-offs where frameworks not only need to optimize the peak or average resource consumption of a pipeline in isolation but optimize its performance-per-resources in the context of complex shared data center infrastructure. We make the case that this environment brings new challenges and opportunities that were previously under-investigated:

\begin{enumerate}
    \item The infrastructure needs to run a large, diverse and ever-changing set of different data processing pipelines across many different inputs. This makes it difficult to have a single heuristic or pipeline configuration that behaves optimally for all of these cases.
    \item Because the pipelines run on shared infrastructure, they compete for resources such as available SSD capacity. Instead of optimizing for peak, they have to make decisions that collectively maximize overall data center utilization.
    \item The diversity of pipelines and runs create the potential to learn from the behavior of one execution and extrapolate to the behavior of others. Similar approaches have been used in query processing~\cite{marcus12neo} and ML compilation \cite{chen2018learning}. However, these techniques operate on highly structured inputs while data processing pipelines run arbitrary code within a dataflow graph and are thus much more difficult to learn.
\end{enumerate}

\noindent We focus on a seemingly innocuous but important problem that emphasizes these challenges: Placement of intermediate data on SSD or HDD. By measuring in-house data processing pipelines running within Google data centers, we observe that a substantial portion of storage resources of these workloads are consumed by reading and writing intermediate data.

In isolation, a pipeline would try to use all available storage resources for this purpose – for example, it would write as much data as possible to SSDs, since SSDs provide higher I/O throughput than HDDs. In a cloud data center, however, each job needs to use SSD resources as cost-effectively as possible since there are usually other jobs that could take advantage of the resources. To run efficiently, data processing frameworks therefore need to decide which data to place on SSD and which data to place on HDD. We find that this problem is challenging to solve – how much an intermediate file benefits from being placed on SSD depends on a number of factors, including the data processing pipeline's code, its inputs, its environment, and the configuration parameters (such as command line flags) used to run the pipeline.

The relationship between these factors and the best placement is non-obvious and poorly understood. In this paper, we analyze Google storage traces to shed light on this relationship. We first propose an analytical model to reason about the economics of deciding when to use SSDs for intermediate files. We then measure a number of real-world in-house data processing pipelines to understand how their I/O behavior varies with changes in their inputs and environments. We find that this relationship is complex and difficult to capture with simple heuristics.

Based on these insights and inspired by prior work in this area \cite{MLSYS2021_82161242}, we describe a vision of how machine learning could be used to optimize intermediate data placement in data processing pipelines. We highlight the similarities and differences to prior work that apply a similar approach to query processing or ML model compilation. Finally, we analyze the underlying prediction problem – which is a form of forecasting \cite{maas2020taxonomy} – and identify challenges and opportunities for tackling this problem.

\section{Data Processing Pipelines}
\label{s:data_processing_pipelines}

\begin{figure}
  \centering
  \includegraphics[width=\linewidth]{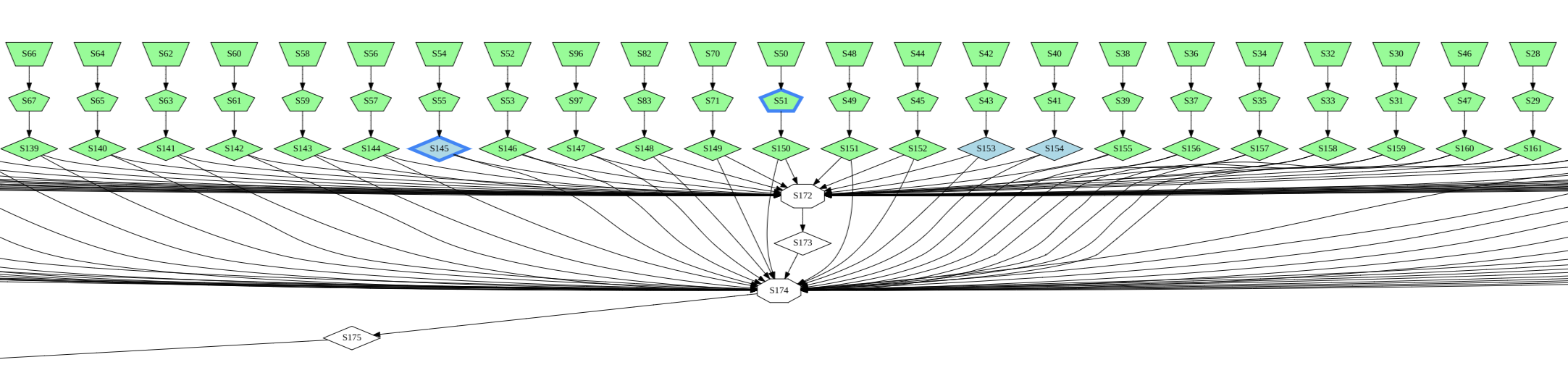}
  \caption{An example of a data processing pipeline. Each box in the figure corresponds a stage. The edges between the stages are the dependencies.}
  \label{fig:pipelines}
\end{figure}

Modern data processing applications apply complex transformations to large data sets. These transformations are represented as data flow graphs. Each node in the data flow graph represents user-defined data transformations, commonly referred to as \emph{stages}. Figure~\ref{fig:pipelines} shows an example data flow graph of one of the jobs in Google's data centers. The nodes represent different stages and the edges represents data flow between stages, sometimes also referred to as data dependencies. We see that most of the stages can run in parallel, but some stages towards the bottom of Figure~\ref{fig:pipelines} depend on inputs from a large number of other stages. To exploit parallizability further, stages are further broken down into \emph{tasks}, which process different \emph{partitions} of the data using one or more operations ~\cite{ananthanarayanan2013effective,ousterhout2017monotasks,ousterhout2013case,ousterhout2015making}. Unlike SQL query engines \cite{marcus12neo}, transformations are not fixed operations but defined as user code written in languages such as Python, Java or C++.

The edges between adjacent stages represent data flow across different transformations. Data dependencies between different stages sometimes require sorting, filtering or grouping of data based on a specified key~\cite{zaharia2012resilient}. In some cases, these operations can be applied to each partition (data for each task) independently. In contrast, some transformations – such as sorting – may require all-to-all communication across all the partitions of the data. This pattern is referred to as a \emph{shuffle}.

\subsection{Temporary Data}

The first stage of a pipeline is responsible for reading data from external sources (e.g., file systems, table storage or streams), while the final stage of the pipeline is responsible for persisting the results of the pipeline. The remaining stages consume intermediate data generated by prior stages, apply data transformations, and generate intermediate data to be consumed by subsequent stages. Communicating intermediate data between stages often incurs network transfers. This is because tasks from successive computation steps may be scheduled on different machines to exploit parallelism. 

Transporting intermediate data between stages typically involves writing temporary intermediate data to storage and reading it back later. This incurs significant I/O overhead as the accesses are often small and random. The purpose of this approach is two-fold: 1) writing data to a distributed storage system allows the data to be transported between tasks running on different machines, and 2) persisting the data protects against task failures which could lead to recomputing the results of one or more stages. In fact, failure of tasks or even cluster nodes is the norm in large-scale deployments, and it is crucial to persist intermediate data for fault tolerance~\cite{huang2017sve,kavulya2010analysis,vishwanath2010characterizing}.

\subsection{Trade-Offs Between HDDs and SSDs}

In a data center setting, the storage layer often consists of a distributed file system~\cite{shvachko2010hdfs,sanjay2003googlegfs}, which can be backed by different storage mediums such as hard drives (HDDs) or solid state drives (SSDs). Data placement is usually decided by the policies of the underlying storage system, but an application may provide hints to the distributed file system to indicate which medium to prefer.

Temporary intermediate data often incurs large amounts of small, random I/O requests, particularly for shuffles~\cite{rao2012sailfish, zhang2018riffle}. HDDs are particularly inefficient at handling those kinds of access patterns. For small, random accesses, the number of available IOPS (I/O Operations Per Second) of a storage medium is generally the limiting factor for system throughput. While HDDs continue to grow in capacity, the available IOPS will not increase accordingly due to the physical limits~\cite{wu2015cloud-diskspin}. Data processing workloads may therefore experience slowdowns if all their temporary files are placed on HDDs. Using multiple disks may increase the available IOPS, but also increases the overall cost significantly.

This problem can be mitigated by placing temporary files on high-performance storage mediums that support large numbers of IOPS, such as SSDs – at a higher cost per GB. However, not all temporary data requires large amounts of IOPS. Data with a low I/O rate may therefore be better placed on HDD, and placing this data also on SSD may lead to sub-optimal usage of this expensive resource.  It is thus important to make judicious decisions about placement of intermediate data, i.e., which data needs to be placed on SSD and which data is best placed on HDD instead.

\subsection{File Placement Decisions}

The decision of where to place data needs to be made at the time of file creation. This is a difficult decision for the distributed file system to make, since features that are available at the file system level (such as file creation time or filename) are not meaningful predictors of the file's access pattern. This is because each data processing pipeline has a variable data flow graph and variable set of transformations. It is thus non-trivial to predict the order in which high I/O density files will be created and accessed.

In cases with limited SSD resources, the optimal policy would be to place all high I/O density files on SSD and the remainder on HDD. However, it is difficult to determine which the high I/O density files are. This becomes even more challenging when multiple workloads are running in parallel and SSD resources need to be shared between these different workloads.

\section{Economics of Storage Devices}
\label{s:econ}

\begin{figure}
  \centering
  \includegraphics[width=\linewidth]{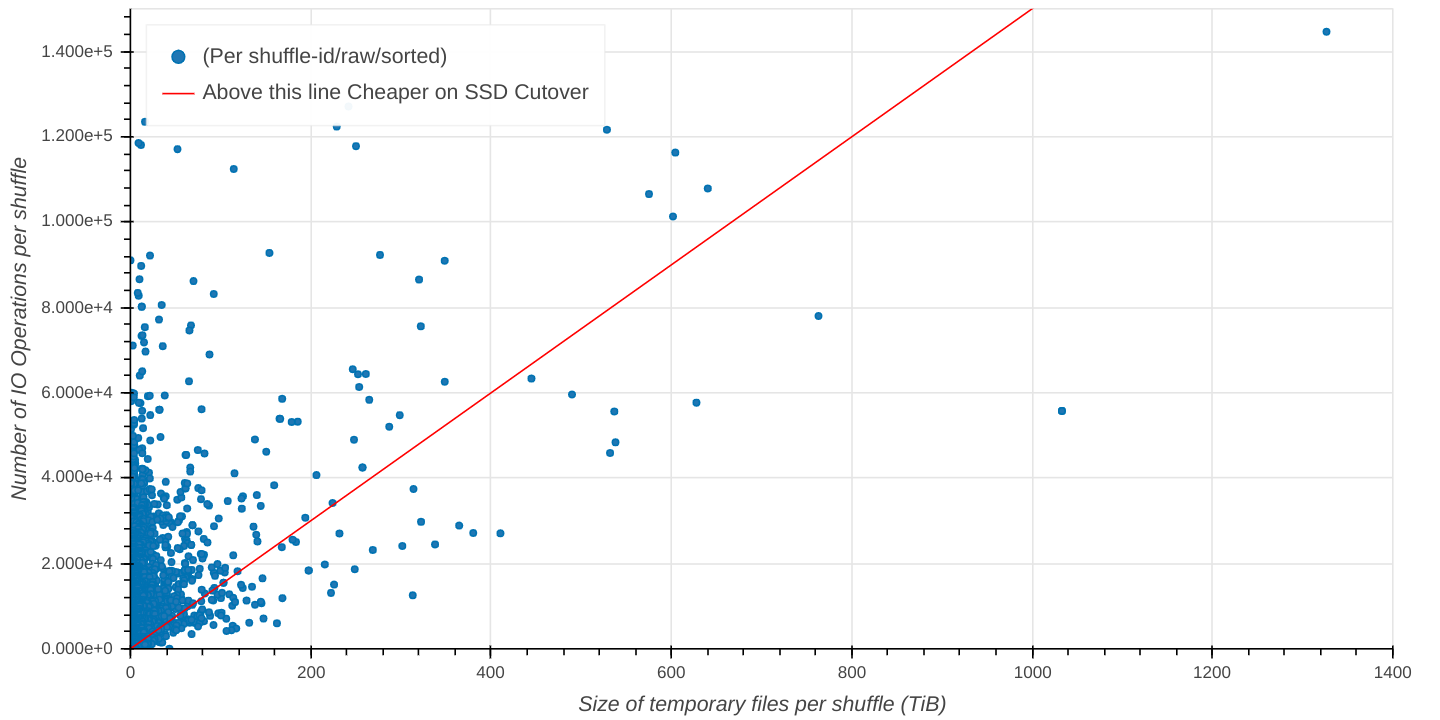}
  \caption{SSD crossover point for shuffles. Each point represents a shuffle operation. Files above the red line exceed the I/O density threshold based on the TCO model.}
  \label{fig:econ_2}
\end{figure}

\begin{figure}
  \centering
  \includegraphics[width=\linewidth]{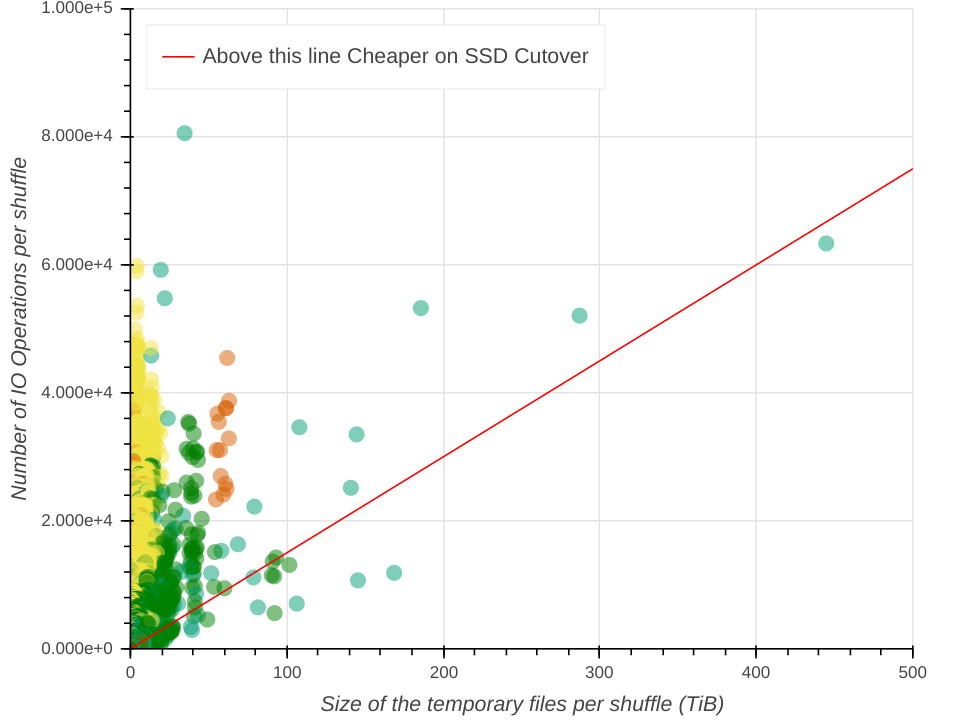}
  \caption{SSD crossover point for different runs of shuffles from the same build. Note that multiple runs from the same build can end up on both sides of the crossover line.} 
  \label{fig:econ_3}
\end{figure}

To support our reasoning about trading off SSD resources between different workloads, we will now introduce an analytical model to capture the economics of the trade-offs between placing data on SSDs and HDDs. As flash technology improves, raw flash cost per byte declines 10-15\% year over year. Commensurately, relative total cost of ownership (TCO) for SSD as compared to HDD declines at a similar rate.

TCO provides a way to consider all the costs associated with purchasing and deploying storage equipment. Typically, TCO for storage devices in distributed clusters is broken down into capital expenditures (CapEx) and operational expenditures (OpEx). CapEx is constrained within the cluster and can include the cost of buying hardware required to deploy a storage solution – the cost of servers, the cost of the storage devices, network hardware, etc. OpEx on the other hand can include the cost of power, depreciation of equipment, maintenance, and repairs.

According to industry standard TCO models, the TCO of one (e.g., 10+ TB) HDD is approximately the same as one TB of SSD. To understand whether the intermediate data associated with a shuffle job would be better placed on SSD or HDD, we can look at whether the required amount of SSD capacity in TB is less than the number of HDDs needed to perform the required amount of I/O.

Our shuffle workloads need both bytes and I/O for the duration of the shuffle. In HDDs, the bytes are cheap, but the IOPS are the dominating factor driving the number of HDDs. For SSDs, there is ample I/O, but the cost per byte is 10$\times$ that of HDDs, and thus the amount of bytes of SSD is the dominating factor. For a particular intermediate file, we can observe whether the amount of space vs. IOPS meets a certain ratio, and use this to decide whether HDD or SSD is a better option. We assume an HDD can perform up to 150 IOPS. An I/O density of more than 150 IOPS per TB is therefore an indicator that the file would be better placed on SSD.

To understand this trade-off in our actual data center-scale deployment, we analyze IOPS/TB of temporary data for various shuffle operations executed by in-house data processing pipelines within a hyperscale data center deployment at Google. Figure~\ref{fig:econ_2} shows the IOPS/TB density for about 1000 shuffle operations executed in one day. Each data point in this graph represents the IOPS density for one shuffle operation. The red line in the graph shows the breakeven point between SSD and HDD. All the shuffles above the red line are cost-effective and should be placed on SSD while all the shuffles below the red line are cheaper to place on HDD.

This calculation does not account for the performance advantage of SSDs. For shuffles above the red line, throughput of the shuffles can also be improved by employing SSD, while also potentially reducing the cost. Figure~\ref{fig:econ_2} shows that for our in-house data processing pipelines, temporary files for about 70\% of the shuffle operations could be placed on SSD without increasing the cost.

This shows that, perhaps somewhat counter-intuitively, it is not always better to place files on SSD but instead there is a 1:2 split in terms of how many files should be placed on HDD vs. SSD from a cost perspective. This demonstrates the importance of making a decision on a per-file basis; placing all files on HDD or placing all files on SSD is insufficient. As we show in Figure~\ref{fig:econ_3}, making a single decision on a per-workload basis is insufficient as well: The trade-off for shuffles across multiple runs of a single binary is similar to the overall distribution.

\section{Problem Description}
\label{s:characterizing}

Prior works often assumed that all the shuffle operations in data processing pipelines are I/O intensive, which suggests those intermediate files should always be placed on SSD. However, Figure~\ref{fig:econ_2} reveals that this is not always the case. Hence, there is a need to optimize storage placement for each operation in the graph of a data processing application individually. 

High I/O rate requirements for data processing pipelines is a known problem~\cite{rao2012sailfish,stuedi2017crail,zhang2018riffle}. There is ample related work which focuses on reducing IOPS by grouping and converting random reads to sequential reads, or by optimizing the graph~\cite{rao2012sailfish,zhang2018riffle,ke2013optimus,yang2019apachenemo}. Others propose using DRAM as much as possible~\cite{condie2010mapreduce,engle2012shark,li2014tachyon}, but in real-world settings, the amount of data is growing much faster than the available memory, which makes it infeasible to keep the data entirely in memory.

In contrast, trading SSD for HDD is still underexplored. While it is possible to retroactively tell whether an operation should have used SSD or HDD, using this approach proactively when making file placement decisions is challenging as it requires predicting I/O density for each data file. I/O density of various operations in a data processing application depends on the structure of the graph of the application as well as the user-level code of each transformation. In addition, there are multiple factors that can introduce variation in the I/O density of temporary files, which make this prediction problem challenging.

\subsection{Sources of I/O Density Variation}

We now discuss different factors that cause variations in the I/O density of temporary files in data processing pipelines. Broadly, these factors fall into three categories: 1) variations across workloads, 2) variations across inputs to the same workload, and 3) variations in the execution environment. We now discuss each in detail.

\vspace{8pt}

\noindent \textbf{Workload: } Each data processing application is different in terms of the transformations it contains. Space and I/O requirements for temporary files associated with each stage are directly impacted by the structure of data flow graph as well as the operations involved in each stage. By measuring real-world data processing pipelines in Google's data centers, we quantify how the different factors affecting the I/O density of temporary files affect these workloads.

Data processing pipelines consist of multiple transformation stages. Temporary files involved in the same stage typically have closely related I/O density as these files are responsible for applying a similar transformations on varying subsets of the input data. However, different stages can have temporary files with highly varied I/O density requirements. This is because the different stages potentially have very different behavior, as they run arbitrary code. Storage space requirement for different transformations can also exhibit high variability. Finally, the same challenges can apply to different versions of the same binaries as code evolves over time.

\vspace{8pt}

\noindent \textbf{Inputs: } The size and content of the input data sets can also influence I/O densities of temporary files. For example, if the input data represents a graph, the connectivity of this graph may affect the density of files generated during shuffle operations.

In addition to the input data sets, configuration parameters, such as command line arguments, can have an impact as well. For example, a pipeline could be configured with a sampling rate that determines which fraction of input records to process. This would introduce additional variability in I/O density.

\vspace{8pt}

\noindent \textbf{Environment:} The load on the distributed clusters can change over time, introducing variability in job completion times. This variability ultimately affects the I/O density of temporary files since the total I/O requirement for the exact same workload will stay the same across multiple runs. One common source of variability in cloud data center environments is diurnal load patterns throughout any given 24-hour period.

\subsection{Predictability of I/O Density}
\label{ss:predictability}

The sources of variability laid out in the previous section make it difficult to predict the I/O density for a particular set of temporary files on the fly. In this section, we will discuss and quantify the challenges that we foresee must be addressed to be able to accurately model I/O densities of temporary files.

\vspace{8pt}

\noindent \textbf{Impact of Workload Changes:} Since a data processing pipeline runs arbitrary code, all pipelines are fundamentally different and it is difficult to reason about a pipeline without ever running it. However, since most data processing pipelines run many times, it is reasonable to look at predictability through the lens of predicting a pipeline's I/O behavior based on prior runs of the same pipeline. However, as we show, this problem in itself is  challenging.

First, pipeline code is not static but gets updated over time. These updates affect both the pipeline operations themselves as well as the underlying software stack, including the data processing framework and runtime libraries. Such updates may change the behavior of the pipeline, and result in different data flow graphs.

Second, the same pipeline code can be executed with different settings. Command line parameters are a good proxy for this effect, as they may affect a wide range of pipeline behavior. In the workloads we experimented with, we found that pipeline-specific command line flags altered the execution of the pipeline and thus the storage requirements for the temporary data files. Note that this focuses on workload-specific command line flags and is different from the execution parameters of data processing frameworks that configure the number of workers and/or tasks for each stage.

\begin{figure}
  \centering
  \includegraphics[width=0.9\linewidth]{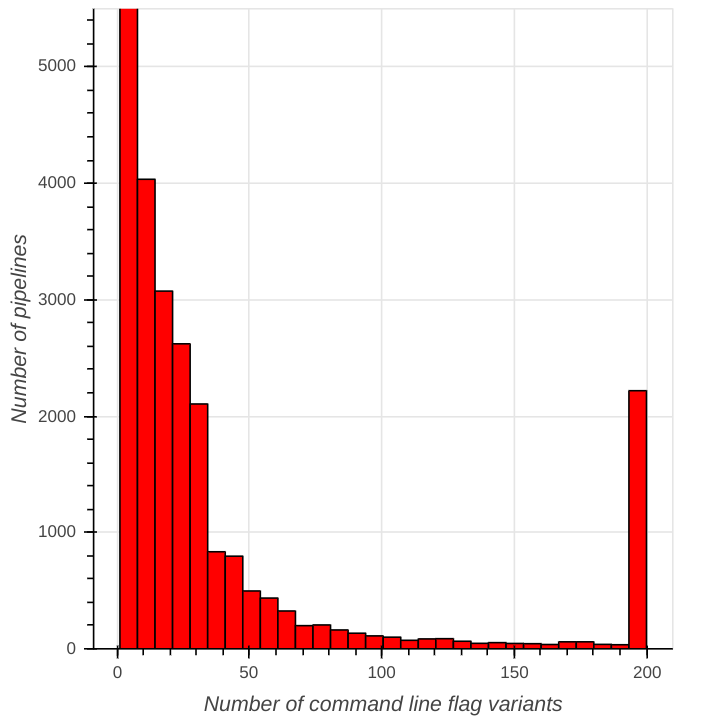}
  \caption{A histogram of the number of different command line flag configurations that we encountered for data processing pipelines at Google. We see that a sizeable fraction of pipelines run with more than one set of command line flags, and a signifciant number of pipelines have more than 200 different configurations.}
  \label{fig:command_line_args}
\end{figure}

As a proxy for the impact of such workload changes, we looked at data processing pipelines at Google and recorded how many different runtime flag configurations they are running with in production. Figure~\ref{fig:command_line_args} shows a histogram of this data. We see that a substantial fraction of workloads has more than one such configuration, and a sizeable fraction of pipelines has 50 or more different variants. This indicates that even if measurements are used to predict I/O density for future runs of the same pipeline, the wide range of different behaviors of the same pipeline needs to be taken into account.

\vspace{8pt}

\begin{figure}
  \centering
  \includegraphics[width=\linewidth]{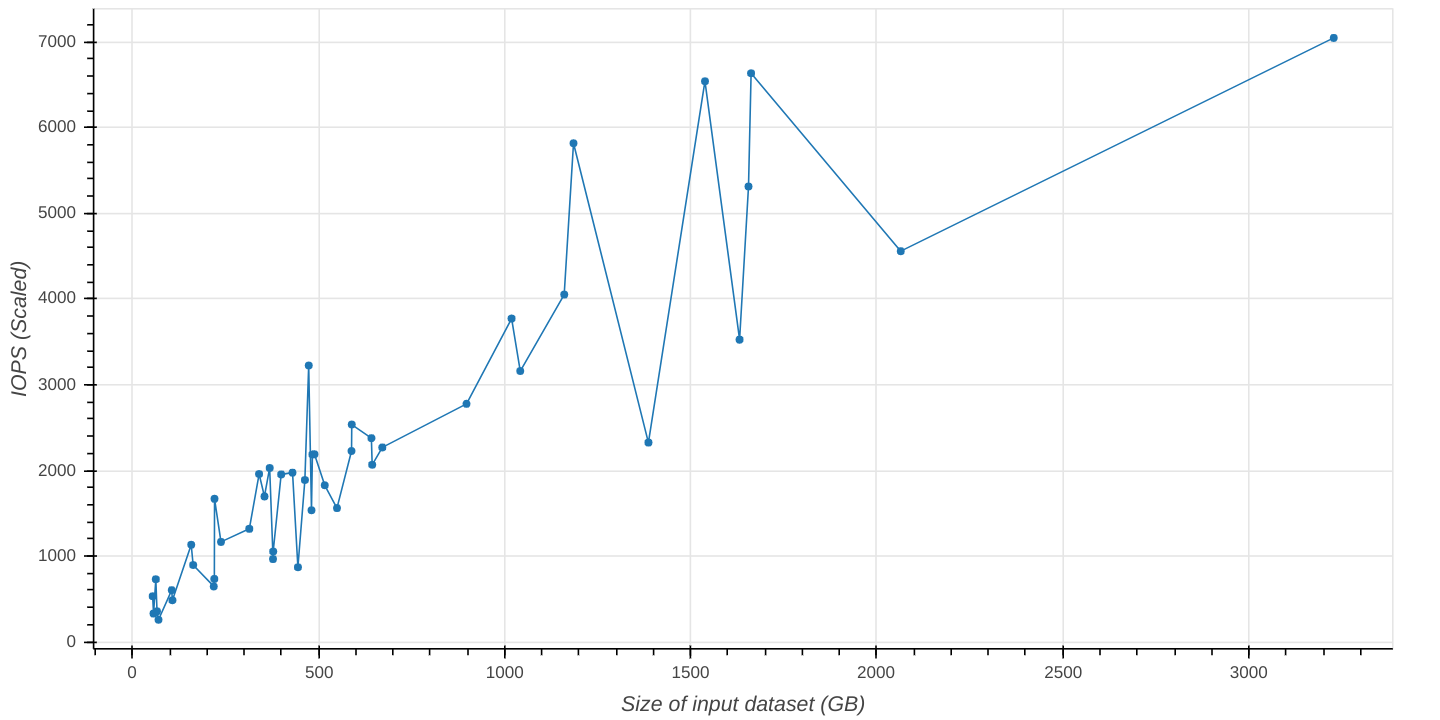}
  \caption{Correlation between I/O rate and input data size for a proprietary data processing pipeline at Google.}
  \label{fig:input_data_size}
\end{figure}

\noindent \textbf{Impact of the Input Size:} Even when running precisely the same data processing pipeline with the same command line arguments, pipeline behavior varies with the input. To quantify this effect, we ran a proprietary Google data processing pipeline against a range of input sizes. Figure~\ref{fig:input_data_size} shows the variation in average IOPS for temporary data of a fixed workload when executed with different inputs. Note that the variation shown in Figure~\ref{fig:input_data_size} is for average IOPS across all the temporary data files. Even though the high level trend in Figure~\ref{fig:input_data_size} shows strong correlation of IOPS with input data size, we observe some variation due to the state of the cluster.

\vspace{8pt}

\noindent \textbf{Impact of the Execution Environment:} Data centers run a large, heterogeneous set of different workloads. Many of these jobs are not data processing applications. The load on the clusters can vary and different workloads may have different priorities. During high load, a lower priority data processing application may get preempted by a higher priority job. In contrast, the same lower priority job may get more resources than the minimum requirements during times of low load.  This variation in the share of resources a data processing application receives introduces variability in job completion times which ultimately affects the I/O density of temporary files, since the total I/O requirement for the workload stays the same. 

\begin{figure}
  \centering
  \includegraphics[width=\linewidth]{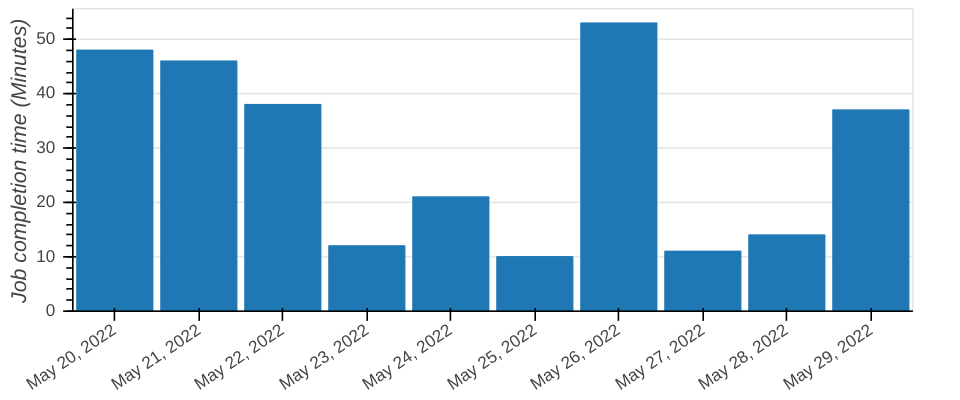}
  \caption{Job completion time variation of a single pipeline using the same data set over multiple invocations.}
  \label{fig:cluster_var}
\end{figure}

To quantify this effect, we looked at logs from production runs of a proprietary data processing pipeline over time. Figure~\ref{fig:cluster_var} shows that the pipeline's execution time exhibited a large variation of completion times across a 10-day period. We execute the  same data processing pipeline (with the same input data) on 10 consecutive days. We see that the job completion time can exhibit variation by up to $5\times$ in this particular scenario. Since the underlying graph and the input data are the same, the total number of IOs incurred by each run of the same workload is fixed. However, the overall IOPS requirement may change because of external factors associated with the state of the cluster. This variation may make predictions of I/O density more challenging, and suggests that such predictions should take the execution environment into account.

\section{Vision: Smart Storage Management}
\label{s:vision}

Data processing pipelines have a large impact on the effective use of hardware resources in cloud data centers. We showed that the optimal placement decision is difficult to predict, due to variations in workloads, inputs and the data center environment (Section~\ref{s:characterizing}). We now describe our vision of future data processing frameworks as an evolution of predicting storage-related properties using machine learning~\cite{MLSYS2021_82161242}. We propose leveraging machine learning to not only predict properties for individual workloads but to manage cross-workload trade-offs across multiple jobs in a data center. We believe that this represents a promising and important research area.

We envision a forecasting framework which optimizes storage placement decisions to increase efficiency of the storage devices in data centers, specifically for data processing workloads. The framework is integrated with distributed data processing frameworks such as Apache Beam or Apache Spark. Its goal is to predict the I/O density for each temporary file at file creation time. Using the predictions about I/O density and availability of each type of storage resource, a data center-level coordination layer could manage the assignment of SSD resources to these different services while taking job priorities and reservations into account. This optimal placement could ultimately maximize aggregate throughput of data processing applications across the cluster while potentially reducing costs. Machine learning could itself be used to identify the best assignment of SSD resources between workloads.

We emphasize that this represents a vision and not a current or planned system at Google. In this section, we discuss the required steps for realizing this vision and identify interesting research challenges that exist in this area.

\subsection{Learning Setup}

\begin{figure}
  \centering
  \includegraphics[width=\linewidth]{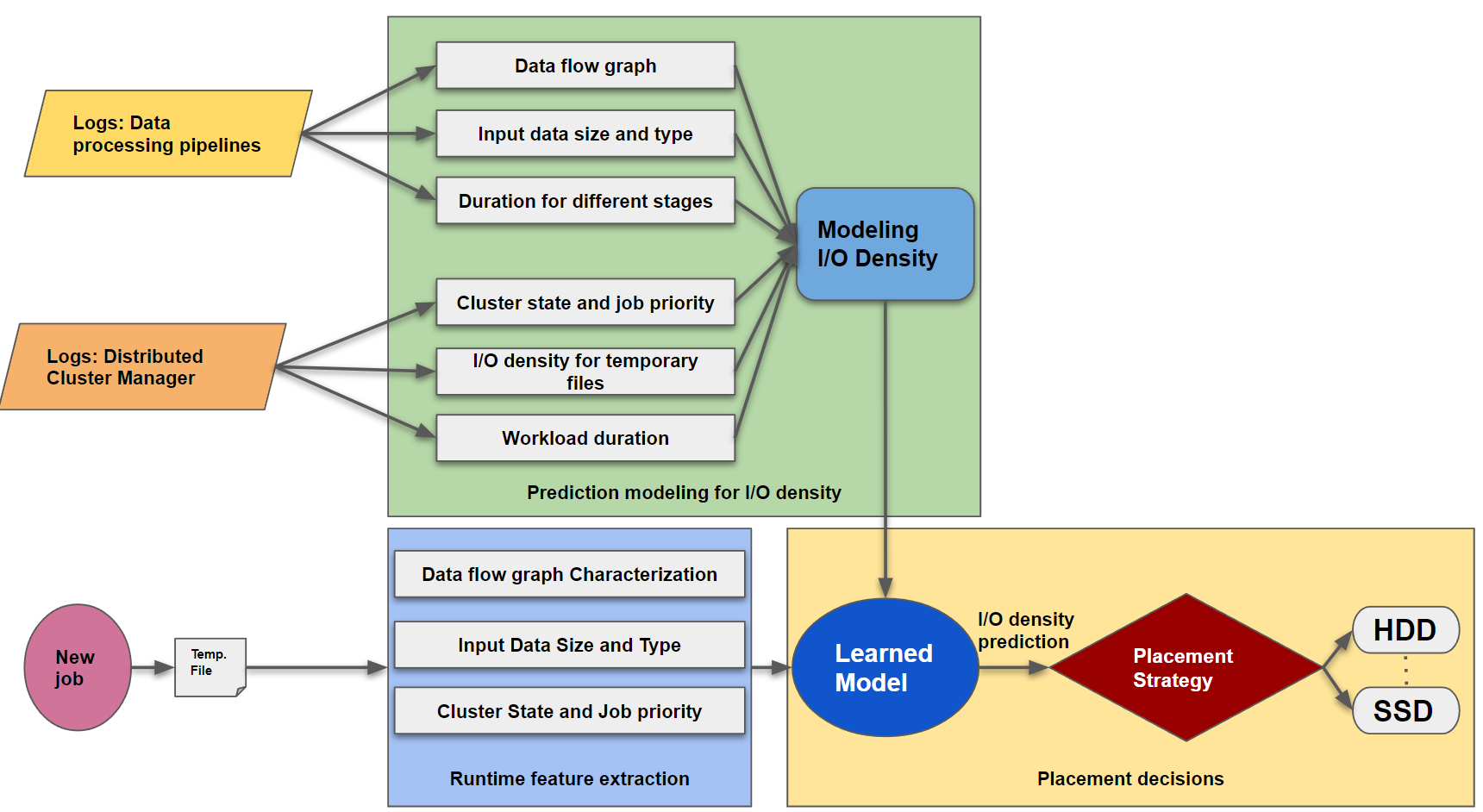}
  \caption{Our envisioned high-level architecture. A model is used to predict the I/O density of newly created temporary files, which is in turn used by a placement strategy that takes the cluster's state into account.}
  \label{fig:sys_design}
\end{figure}

Forecasting models can leverage signals from the underlying data flow graph, input data, and distributed cluster to make placement decisions. Figure~\ref{fig:sys_design} shows the high-level proposed predictor design. The forecasting framework can leverage various signals from historical logs associated with data processing pipelines executed in the data center. Signals from these logs, such as data flow graphs, I/O density of temporary files, job priorities, etc. can be extracted and employed to model and predict I/O density for temporary files created by future workloads. The model targets both new workloads as well as repeated runs of previously observed workloads.

Once we are able to characterize files by learning across multiple dimensions, whenever a new files is created by a data processing application, various associated features about data processing application as well as the state of the cluster are passed to the learned model to predict anticipated I/O density for each file. These predictions are then employed to make placement decisions to optimize SSD as well as HDD usage. The following subsections discuss the three main challenging components in the design of the forecasting framework: 1) Prediction and modeling of I/O density of newly created temporary files, 2) Designing placement policies, and 3) Integration of the I/O density forecasting framework with existing data processing frameworks.

\subsection{Features and Models for I/O Density}

For predicting I/O density of temporary files, our analysis reveals that for each data processing workload, it is important to characterize input data, the underlying data flow graph, and the behavior of the distributed cluster on which the job is being executed. We discuss how to model these features in the following paragraphs.

\vspace{8pt}

\noindent \textbf{Signals from input data: }We characterize each workload based on the program binary as well as the underlying graph. For a fixed program binary, the underlying graph can change with variations in input parameters from the user at runtime. We can use a combination of program binary as well as underlying data flow graph to identify unique workloads. We observe that for a given workload, size as well as type of input data exhibit close correlation with the I/O density of the temporary files involved in different stages of the data processing application (see Section~\ref{ss:predictability} for more details).

There is a need to design and integrate APIs with data processing frameworks to employ statistics related to input data in storage placement decisions. These statistics can facilitate I/O density predictions which can ultimately be utilized to improve data placement policies. For structured input data, there may also be opportunities to learn on structural properties of the data rather than just the input size – for example, the degree of connectivity of a graph representation of the data.

\vspace{7pt}

\noindent \textbf{Modeling the data flow graph: } Each data processing job is defined by the structure of the data flow graph as well as the compute complexity of each stage in the graph. These are important parameters of a data processing application which can dictate the  I/O requirements of the job for each stage. Existing APIs can provide unstructured information about these parameters of the application, for example, graph structure and the code associated with each node. Learning on this unstructured data is challenging, but neural networks have made great progress in both areas, including graph neural networks (GNNs) that can capture graph properties~\cite{gnn-scarselli2008graph} and a range of models for program code \cite{allamanis2018survey}.

\vspace{7pt}

\noindent \textbf{Approximating behavior of hardware: }In addition to information about the data processing application, estimating behavior of the underlying distributed cluster on which the job is scheduled is also important. The behavior of distributed clusters depends on multiple factors, such as the amount of load at a given time, priority of the job, amount of resources requested, etc.

Each public cluster might respond differently to these parameters. However, distributed clusters are typically long running and usually keep logs for both long-term as well as short-term history. Learning-based techniques can be employed to process logs and extract correlations between job completion time and the parameters which can alter the job completion time, e.g., current load on the cluster, priority of the job, etc. (see Section~\ref{ss:predictability} for more details). 

Another way to model the runtime variations due to the underlying hardware is by modeling I/O density as well as job completion time as probability distributions which may require additional data similar to what is shown in Figure~\ref{fig:cluster_var}.
This kind of data may not be available for all workloads.

\vspace{8pt}

\noindent \textbf{Ground truth labels: }While the previous paragraphs capture features needed by the model, we also need \emph{labels} to train against – the ground truth that determines what I/O density the features correspond to.
Automating extraction of information about disk accesses (e.g., size of reads, number of I/O operations etc.) is non-trivial.
Some of this data may be extracted from existing profiling infrastructure used to debug and analyze storage systems (e.g., similar to the traces in \cite{MLSYS2021_82161242}).
Other data may benefit from additional hooks to reliably collect and aggregate it during pipeline execution.

\vspace{8pt}

\noindent \textbf{Suitable Model Architectures: }The model needs to balance two goals: 1) integrate the different signals described in the previous paragraph, including free-form program code and the data flow graph structure, while 2) being compact enough that it can be run within the context of a data processing pipeline without major overheads. One approach to achieve this could be to train a large neural network that can combine the data, and then precompute some of the components, such as the parts ingesting the code and graph structure (as they are static). It may also be possible to distill the model into a cheaper approach such as decision trees or random forests. Finally, models can become cheaper by learning a small and specialized model for every place where a temporary file is created, rather than a large model that handles all of them together.

\subsection{Designing Placement Policies}

Once I/O densities have been predicted, these predictions can be communicated to a central service that makes the final decision for SSD/HDD placement for all workloads in the cluster.

Based on our analysis in Section~\ref{s:econ}, all the temporary files with IOPS/TB of data is greater than 150 can be placed on SSD without increasing the cost and can potentially reduce job completion times. However, the amount of SSD may be limited and cannot host all the files that can potentially benefit from being placed on SSDs. 

To maximize efficiency of SSDs, the temporary files which save the largest amount of IOPS per TB of space used must be placed on SSD. In scenarios where SSD is limited, a scoring-based technique can be implemented which assigns a score to each temporary files by computing the ratio of expected IOPS and the size of the file. To employ this policy, the files placed on SSD must be sorted based on the score. When a new file is created, its expected score is calculated using IOPS and file size predictions. Based on the predicted score, if replacing a subset of existing files placed on SSD improves SSD efficiency, the file can be placed on SSD, moving some older files to HDD. Otherwise the file should simply be placed on HDD.

\subsection{System Integration}

Making all the required features available for prediction at all times is challenging.
Typically, data processing applications are executed in distributed clusters across multiple processes on different workers. While using multiple processes speeds up the execution of the application, it makes it difficult to keep a coherent view of all the parameters which can effect the I/O density of temporary files as these parameters will be distributed across different workers as well. Additionally, the state of the distributed cluster (which is often dynamic) can vary and affect I/O density of the temporary files. This necessitates a framework that assembles the features for I/O density predictions during the execution of the data processing application and ensures that all features are available for placement decisions whenever a new temporary file is being created.

We propose to address this through a general library that intercepts file creation and automatically collects the required features. This means that there is a central portion of every process that can handle predictions. The outcome of the prediction is then added to the metadata that is attached to every file request, which makes it visible to the central distributed file system that can make a system-wide placement decision, trading off different workloads.

\section{Related Work}
\label{s:related_works}

Prior works focused on improving efficiency of data processing pipelines by optimizing across multiple dimensions. To reduce IOPS frequency for intermediate data, specialized tools have been proposed for improving data locality~\cite{zhang2018riffle}. Further, these tools aggregate storage accesses (reads, writes, etc.) to piggyback on IOPS~\cite{zhang2018riffle, rao2012sailfish}. Apache Nemo~\cite{yang2019apachenemo} proposes a specialized intermediate storage representation to minimize IOPS for shuffle operations in data processing pipelines. Additionally, there has been a focus on improving computational efficiency as well by applying various techniques to optimize the data flow graphs~\cite{yang2019apachenemo,rao2012sailfish}. There are also tools which implement techniques to increase parallelizability of compute operations in data processing pipelines to utilize high performance clusters available in cloud deployments~\cite{isard2007dryad,ke2013optimus}.

Most of the above prior works focus on improving resource efficiency for data processing pipelines when deployed in isolation and ignore scenarios when resources are shared among multiple pipelines (e.g., in cloud-based deployments). In contrast, we identify challenges involved in optimizing storage efficiency of data processing pipelines for cloud resources from a data center perspective, considering resource usage for multiple pipelines together. 

Recent research has demonstrated the effectiveness of machine learning for various efficiency objectives in cloud deployments. 3Sigma~\cite{park20183sigma} implements a learning-based job scheduler to increase throughput for various clusters. Quasar~\cite{delimitrou2014quasar} and ResourceCentral~\cite{cortez2017resource} have employed ML to significantly improve scheduling decisions by employing cheap models. Seer~\cite{gan2019seer} is able to predict QoS violations before they occur, using historical trace data.

Existing ML-based implementations, such as 3Sigma, use application features (e.g., user id, program name, etc.) for predictions. Such application features are readily accessible for cluster schedulers. However, exploiting these features for storage devices is challenging. Sometimes, each feature may have assigned multiple values over the lifetime of a given request, since a given request may pass through multiple hosts. For example, a user may request some data using a front-end engine. The front-end passes the request to a back-end database server, which ultimately fetches the requested data from a shared storage device deployed in a cluster. 

Some of the prior works~\cite{kirilin2019rl,song2020learning,liu2020imitation} implement learning-based caching policies to improve efficiency of storage resources by increasing cache hitrate. These works are closely related to the vision proposed in this paper. However, it is important to note that the existing systems focus on improving heuristic-based caching algorithms while using readily available conventional features. However, this paper envisions ML to extract new application-level features, enabling the design of novel storage management algorithms, specifically tailored for data processing pipelines.

\section{Conclusion}

We believe that learning from prior data processing pipeline executions and applying the results to new pipelines and inputs is an important opportunity for real-world deployments of data processing pipelines in cloud data centers. It is understudied in the current literature. We hope to make the cloud computing community aware of this new and promising research area.

\bibliographystyle{ACM-Reference-Format}
\bibliography{refs}


\begin{thebibliography}{37}


\ifx \showCODEN    \undefined \def \showCODEN     #1{\unskip}     \fi
\ifx \showDOI      \undefined \def \showDOI       #1{#1}\fi
\ifx \showISBNx    \undefined \def \showISBNx     #1{\unskip}     \fi
\ifx \showISBNxiii \undefined \def \showISBNxiii  #1{\unskip}     \fi
\ifx \showISSN     \undefined \def \showISSN      #1{\unskip}     \fi
\ifx \showLCCN     \undefined \def \showLCCN      #1{\unskip}     \fi
\ifx \shownote     \undefined \def \shownote      #1{#1}          \fi
\ifx \showarticletitle \undefined \def \showarticletitle #1{#1}   \fi
\ifx \showURL      \undefined \def \showURL       {\relax}        \fi
\providecommand\bibfield[2]{#2}
\providecommand\bibinfo[2]{#2}
\providecommand\natexlab[1]{#1}
\providecommand\showeprint[2][]{arXiv:#2}

\bibitem[apa(2020)]%
        {apache_beam}
 \bibinfo{year}{2020}\natexlab{}.
\newblock \bibinfo{title}{Apache Beam: An advanced unified programming model.}
\newblock \bibinfo{howpublished}{\url{https://beam.apache.org/}}.
\newblock


\bibitem[Allamanis et~al\mbox{.}(2018)]%
        {allamanis2018survey}
\bibfield{author}{\bibinfo{person}{Miltiadis Allamanis},
  \bibinfo{person}{Earl~T Barr}, \bibinfo{person}{Premkumar Devanbu}, {and}
  \bibinfo{person}{Charles Sutton}.} \bibinfo{year}{2018}\natexlab{}.
\newblock \showarticletitle{A survey of machine learning for big code and
  naturalness}.
\newblock \bibinfo{journal}{\emph{ACM Computing Surveys (CSUR)}}
  \bibinfo{volume}{51}, \bibinfo{number}{4} (\bibinfo{year}{2018}),
  \bibinfo{pages}{81}.
\newblock


\bibitem[Ananthanarayanan et~al\mbox{.}(2013)]%
        {ananthanarayanan2013effective}
\bibfield{author}{\bibinfo{person}{Ganesh Ananthanarayanan},
  \bibinfo{person}{Ali Ghodsi}, \bibinfo{person}{Scott Shenker}, {and}
  \bibinfo{person}{Ion Stoica}.} \bibinfo{year}{2013}\natexlab{}.
\newblock \showarticletitle{Effective straggler mitigation: Attack of the
  clones}. In \bibinfo{booktitle}{\emph{10th USENIX Symposium on Networked
  Systems Design and Implementation (NSDI 13)}}. \bibinfo{pages}{185--198}.
\newblock


\bibitem[Chen et~al\mbox{.}(2018)]%
        {chen2018learning}
\bibfield{author}{\bibinfo{person}{Tianqi Chen}, \bibinfo{person}{Lianmin
  Zheng}, \bibinfo{person}{Eddie Yan}, \bibinfo{person}{Ziheng Jiang},
  \bibinfo{person}{Thierry Moreau}, \bibinfo{person}{Luis Ceze},
  \bibinfo{person}{Carlos Guestrin}, {and} \bibinfo{person}{Arvind
  Krishnamurthy}.} \bibinfo{year}{2018}\natexlab{}.
\newblock \showarticletitle{Learning to optimize tensor programs}.
\newblock \bibinfo{journal}{\emph{Advances in Neural Information Processing
  Systems}}  \bibinfo{volume}{31} (\bibinfo{year}{2018}).
\newblock


\bibitem[Condie et~al\mbox{.}(2010)]%
        {condie2010mapreduce}
\bibfield{author}{\bibinfo{person}{Tyson Condie}, \bibinfo{person}{Neil
  Conway}, \bibinfo{person}{Peter Alvaro}, \bibinfo{person}{Joseph~M
  Hellerstein}, \bibinfo{person}{Khaled Elmeleegy}, {and}
  \bibinfo{person}{Russell Sears}.} \bibinfo{year}{2010}\natexlab{}.
\newblock \showarticletitle{MapReduce online.}. In
  \bibinfo{booktitle}{\emph{Nsdi}}, Vol.~\bibinfo{volume}{10}.
  \bibinfo{pages}{20}.
\newblock


\bibitem[Cortez et~al\mbox{.}(2017)]%
        {cortez2017resource}
\bibfield{author}{\bibinfo{person}{Eli Cortez}, \bibinfo{person}{Anand Bonde},
  \bibinfo{person}{Alexandre Muzio}, \bibinfo{person}{Mark Russinovich},
  \bibinfo{person}{Marcus Fontoura}, {and} \bibinfo{person}{Ricardo
  Bianchini}.} \bibinfo{year}{2017}\natexlab{}.
\newblock \showarticletitle{Resource central: Understanding and predicting
  workloads for improved resource management in large cloud platforms}. In
  \bibinfo{booktitle}{\emph{Proceedings of the 26th Symposium on Operating
  Systems Principles}}. \bibinfo{pages}{153--167}.
\newblock


\bibitem[Delimitrou and Kozyrakis(2014)]%
        {delimitrou2014quasar}
\bibfield{author}{\bibinfo{person}{Christina Delimitrou} {and}
  \bibinfo{person}{Christos Kozyrakis}.} \bibinfo{year}{2014}\natexlab{}.
\newblock \showarticletitle{Quasar: Resource-efficient and qos-aware cluster
  management}.
\newblock \bibinfo{journal}{\emph{ACM SIGPLAN Notices}} \bibinfo{volume}{49},
  \bibinfo{number}{4} (\bibinfo{year}{2014}), \bibinfo{pages}{127--144}.
\newblock


\bibitem[Engle et~al\mbox{.}(2012)]%
        {engle2012shark}
\bibfield{author}{\bibinfo{person}{Cliff Engle}, \bibinfo{person}{Antonio
  Lupher}, \bibinfo{person}{Reynold Xin}, \bibinfo{person}{Matei Zaharia},
  \bibinfo{person}{Michael~J Franklin}, \bibinfo{person}{Scott Shenker}, {and}
  \bibinfo{person}{Ion Stoica}.} \bibinfo{year}{2012}\natexlab{}.
\newblock \showarticletitle{Shark: fast data analysis using coarse-grained
  distributed memory}. In \bibinfo{booktitle}{\emph{Proceedings of the 2012 ACM
  SIGMOD International Conference on Management of Data}}.
  \bibinfo{pages}{689--692}.
\newblock


\bibitem[Feng and Chen(2016)]%
        {feng2016optimization}
\bibfield{author}{\bibinfo{person}{Yunping Feng} {and} \bibinfo{person}{Haopeng
  Chen}.} \bibinfo{year}{2016}\natexlab{}.
\newblock \showarticletitle{Optimization of spark storage solutions}. In
  \bibinfo{booktitle}{\emph{2016 International Conference on Progress in
  Informatics and Computing (PIC)}}. IEEE, \bibinfo{pages}{473--478}.
\newblock


\bibitem[Gan et~al\mbox{.}(2019)]%
        {gan2019seer}
\bibfield{author}{\bibinfo{person}{Yu Gan}, \bibinfo{person}{Yanqi Zhang},
  \bibinfo{person}{Kelvin Hu}, \bibinfo{person}{Dailun Cheng},
  \bibinfo{person}{Yuan He}, \bibinfo{person}{Meghna Pancholi}, {and}
  \bibinfo{person}{Christina Delimitrou}.} \bibinfo{year}{2019}\natexlab{}.
\newblock \showarticletitle{Seer: Leveraging big data to navigate the
  complexity of performance debugging in cloud microservices}. In
  \bibinfo{booktitle}{\emph{Proceedings of the twenty-fourth international
  conference on architectural support for programming languages and operating
  systems}}. \bibinfo{pages}{19--33}.
\newblock


\bibitem[Huang et~al\mbox{.}(2017)]%
        {huang2017sve}
\bibfield{author}{\bibinfo{person}{Qi Huang}, \bibinfo{person}{Petchean Ang},
  \bibinfo{person}{Peter Knowles}, \bibinfo{person}{Tomasz Nykiel},
  \bibinfo{person}{Iaroslav Tverdokhlib}, \bibinfo{person}{Amit Yajurvedi},
  \bibinfo{person}{Paul Dapolito~IV}, \bibinfo{person}{Xifan Yan},
  \bibinfo{person}{Maxim Bykov}, \bibinfo{person}{Chuen Liang},
  {et~al\mbox{.}}} \bibinfo{year}{2017}\natexlab{}.
\newblock \showarticletitle{SVE: Distributed video processing at Facebook
  scale}. In \bibinfo{booktitle}{\emph{Proceedings of the 26th Symposium on
  Operating Systems Principles}}. \bibinfo{pages}{87--103}.
\newblock


\bibitem[Isard et~al\mbox{.}(2007)]%
        {isard2007dryad}
\bibfield{author}{\bibinfo{person}{Michael Isard}, \bibinfo{person}{Mihai
  Budiu}, \bibinfo{person}{Yuan Yu}, \bibinfo{person}{Andrew Birrell}, {and}
  \bibinfo{person}{Dennis Fetterly}.} \bibinfo{year}{2007}\natexlab{}.
\newblock \showarticletitle{Dryad: distributed data-parallel programs from
  sequential building blocks}. In \bibinfo{booktitle}{\emph{Proceedings of the
  2nd ACM SIGOPS/EuroSys European Conference on Computer Systems 2007}}.
  \bibinfo{pages}{59--72}.
\newblock


\bibitem[Kavulya et~al\mbox{.}(2010)]%
        {kavulya2010analysis}
\bibfield{author}{\bibinfo{person}{Soila Kavulya}, \bibinfo{person}{Jiaqi Tan},
  \bibinfo{person}{Rajeev Gandhi}, {and} \bibinfo{person}{Priya Narasimhan}.}
  \bibinfo{year}{2010}\natexlab{}.
\newblock \showarticletitle{An analysis of traces from a production mapreduce
  cluster}. In \bibinfo{booktitle}{\emph{2010 10th IEEE/ACM International
  Conference on Cluster, Cloud and Grid Computing}}. IEEE,
  \bibinfo{pages}{94--103}.
\newblock


\bibitem[Ke et~al\mbox{.}(2013)]%
        {ke2013optimus}
\bibfield{author}{\bibinfo{person}{Qifa Ke}, \bibinfo{person}{Michael Isard},
  {and} \bibinfo{person}{Yuan Yu}.} \bibinfo{year}{2013}\natexlab{}.
\newblock \showarticletitle{Optimus: a dynamic rewriting framework for
  data-parallel execution plans}. In \bibinfo{booktitle}{\emph{Proceedings of
  the 8th ACM European Conference on Computer Systems}}.
  \bibinfo{pages}{15--28}.
\newblock


\bibitem[Kirilin et~al\mbox{.}(2019)]%
        {kirilin2019rl}
\bibfield{author}{\bibinfo{person}{Vadim Kirilin}, \bibinfo{person}{Aditya
  Sundarrajan}, \bibinfo{person}{Sergey Gorinsky}, {and}
  \bibinfo{person}{Ramesh~K Sitaraman}.} \bibinfo{year}{2019}\natexlab{}.
\newblock \showarticletitle{Rl-cache: Learning-based cache admission for
  content delivery}. In \bibinfo{booktitle}{\emph{Proceedings of the 2019
  Workshop on Network Meets AI \& ML}}. \bibinfo{pages}{57--63}.
\newblock


\bibitem[Li et~al\mbox{.}(2014)]%
        {li2014tachyon}
\bibfield{author}{\bibinfo{person}{Haoyuan Li}, \bibinfo{person}{Ali Ghodsi},
  \bibinfo{person}{Matei Zaharia}, \bibinfo{person}{Scott Shenker}, {and}
  \bibinfo{person}{Ion Stoica}.} \bibinfo{year}{2014}\natexlab{}.
\newblock \showarticletitle{Tachyon: Reliable, memory speed storage for cluster
  computing frameworks}. In \bibinfo{booktitle}{\emph{Proceedings of the ACM
  Symposium on Cloud Computing}}. \bibinfo{pages}{1--15}.
\newblock


\bibitem[Liu et~al\mbox{.}(2020)]%
        {liu2020imitation}
\bibfield{author}{\bibinfo{person}{Evan Liu}, \bibinfo{person}{Milad Hashemi},
  \bibinfo{person}{Kevin Swersky}, \bibinfo{person}{Parthasarathy Ranganathan},
  {and} \bibinfo{person}{Junwhan Ahn}.} \bibinfo{year}{2020}\natexlab{}.
\newblock \showarticletitle{An imitation learning approach for cache
  replacement}. In \bibinfo{booktitle}{\emph{International Conference on
  Machine Learning}}. PMLR, \bibinfo{pages}{6237--6247}.
\newblock


\bibitem[Maas(2020)]%
        {maas2020taxonomy}
\bibfield{author}{\bibinfo{person}{Martin Maas}.}
  \bibinfo{year}{2020}\natexlab{}.
\newblock \showarticletitle{A taxonomy of ML for systems problems}.
\newblock \bibinfo{journal}{\emph{IEEE Micro}}  \bibinfo{volume}{40}
  (\bibinfo{year}{2020}), \bibinfo{pages}{8--16}.
\newblock


\bibitem[Marcus et~al\mbox{.}(2019)]%
        {marcus12neo}
\bibfield{author}{\bibinfo{person}{Ryan Marcus}, \bibinfo{person}{Parimarjan
  Negi}, \bibinfo{person}{Hongzi Mao}, \bibinfo{person}{Chi Zhang},
  \bibinfo{person}{Mohammad Alizadeh}, \bibinfo{person}{Tim Kraska},
  \bibinfo{person}{Olga Papaemmanouil}, {and} \bibinfo{person}{Nesime
  Tatbul23}.} \bibinfo{year}{2019}\natexlab{}.
\newblock \showarticletitle{Neo: A Learned Query Optimizer}.
\newblock \bibinfo{journal}{\emph{Proceedings of the VLDB Endowment}}
  \bibinfo{volume}{12} (\bibinfo{year}{2019}).
\newblock


\bibitem[Ousterhout et~al\mbox{.}(2017)]%
        {ousterhout2017monotasks}
\bibfield{author}{\bibinfo{person}{Kay Ousterhout},
  \bibinfo{person}{Christopher Canel}, \bibinfo{person}{Sylvia Ratnasamy},
  {and} \bibinfo{person}{Scott Shenker}.} \bibinfo{year}{2017}\natexlab{}.
\newblock \showarticletitle{Monotasks: Architecting for performance clarity in
  data analytics frameworks}. In \bibinfo{booktitle}{\emph{Proceedings of the
  26th Symposium on Operating Systems Principles}}. \bibinfo{pages}{184--200}.
\newblock


\bibitem[Ousterhout et~al\mbox{.}(2013)]%
        {ousterhout2013case}
\bibfield{author}{\bibinfo{person}{Kay Ousterhout}, \bibinfo{person}{Aurojit
  Panda}, \bibinfo{person}{Joshua Rosen}, \bibinfo{person}{Shivaram
  Venkataraman}, \bibinfo{person}{Reynold Xin}, \bibinfo{person}{Sylvia
  Ratnasamy}, \bibinfo{person}{Scott Shenker}, {and} \bibinfo{person}{Ion
  Stoica}.} \bibinfo{year}{2013}\natexlab{}.
\newblock \showarticletitle{The case for tiny tasks in compute clusters}. In
  \bibinfo{booktitle}{\emph{14th Workshop on Hot Topics in Operating Systems
  (HotOS XIV)}}.
\newblock


\bibitem[Ousterhout et~al\mbox{.}(2015)]%
        {ousterhout2015making}
\bibfield{author}{\bibinfo{person}{Kay Ousterhout}, \bibinfo{person}{Ryan
  Rasti}, \bibinfo{person}{Sylvia Ratnasamy}, \bibinfo{person}{Scott Shenker},
  {and} \bibinfo{person}{Byung-Gon Chun}.} \bibinfo{year}{2015}\natexlab{}.
\newblock \showarticletitle{Making sense of performance in data analytics
  frameworks}. In \bibinfo{booktitle}{\emph{12th USENIX Symposium on Networked
  Systems Design and Implementation (NSDI 15)}}. \bibinfo{pages}{293--307}.
\newblock


\bibitem[Park et~al\mbox{.}(2018)]%
        {park20183sigma}
\bibfield{author}{\bibinfo{person}{Jun~Woo Park}, \bibinfo{person}{Alexey
  Tumanov}, \bibinfo{person}{Angela Jiang}, \bibinfo{person}{Michael~A Kozuch},
  {and} \bibinfo{person}{Gregory~R Ganger}.} \bibinfo{year}{2018}\natexlab{}.
\newblock \showarticletitle{3sigma: distribution-based cluster scheduling for
  runtime uncertainty}. In \bibinfo{booktitle}{\emph{Proceedings of the
  Thirteenth EuroSys Conference}}. \bibinfo{pages}{1--17}.
\newblock


\bibitem[Rao et~al\mbox{.}(2012)]%
        {rao2012sailfish}
\bibfield{author}{\bibinfo{person}{Sriram Rao}, \bibinfo{person}{Raghu
  Ramakrishnan}, \bibinfo{person}{Adam Silberstein}, \bibinfo{person}{Mike
  Ovsiannikov}, {and} \bibinfo{person}{Damian Reeves}.}
  \bibinfo{year}{2012}\natexlab{}.
\newblock \showarticletitle{Sailfish: A framework for large scale data
  processing}. In \bibinfo{booktitle}{\emph{Proceedings of the Third ACM
  Symposium on Cloud Computing}}. \bibinfo{pages}{1--14}.
\newblock


\bibitem[Rasmussen et~al\mbox{.}(2012)]%
        {rasmussen2012themis}
\bibfield{author}{\bibinfo{person}{Alexander Rasmussen},
  \bibinfo{person}{Vinh~The Lam}, \bibinfo{person}{Michael Conley},
  \bibinfo{person}{George Porter}, \bibinfo{person}{Rishi Kapoor}, {and}
  \bibinfo{person}{Amin Vahdat}.} \bibinfo{year}{2012}\natexlab{}.
\newblock \showarticletitle{Themis: an i/o-efficient mapreduce}. In
  \bibinfo{booktitle}{\emph{Proceedings of the Third ACM Symposium on Cloud
  Computing}}. \bibinfo{pages}{1--14}.
\newblock


\bibitem[Sanjay(2003)]%
        {sanjay2003googlegfs}
\bibfield{author}{\bibinfo{person}{Ghemawat Sanjay}.}
  \bibinfo{year}{2003}\natexlab{}.
\newblock \showarticletitle{The Google file system}. In
  \bibinfo{booktitle}{\emph{SOSP'03: Proceedings of the nineteenth ACM
  symposium on Operating systems principles}}. ACM Press,
  \bibinfo{pages}{29--43}.
\newblock


\bibitem[Scarselli et~al\mbox{.}(2008)]%
        {gnn-scarselli2008graph}
\bibfield{author}{\bibinfo{person}{Franco Scarselli}, \bibinfo{person}{Marco
  Gori}, \bibinfo{person}{Ah~Chung Tsoi}, \bibinfo{person}{Markus
  Hagenbuchner}, {and} \bibinfo{person}{Gabriele Monfardini}.}
  \bibinfo{year}{2008}\natexlab{}.
\newblock \showarticletitle{The graph neural network model}.
\newblock \bibinfo{journal}{\emph{IEEE transactions on neural networks}}
  \bibinfo{volume}{20}, \bibinfo{number}{1} (\bibinfo{year}{2008}),
  \bibinfo{pages}{61--80}.
\newblock


\bibitem[Shvachko et~al\mbox{.}(2010)]%
        {shvachko2010hdfs}
\bibfield{author}{\bibinfo{person}{Konstantin Shvachko},
  \bibinfo{person}{Hairong Kuang}, \bibinfo{person}{Sanjay Radia}, {and}
  \bibinfo{person}{Robert Chansler}.} \bibinfo{year}{2010}\natexlab{}.
\newblock \showarticletitle{The hadoop distributed file system}. In
  \bibinfo{booktitle}{\emph{2010 IEEE 26th symposium on mass storage systems
  and technologies (MSST)}}. Ieee, \bibinfo{pages}{1--10}.
\newblock


\bibitem[Song et~al\mbox{.}(2020)]%
        {song2020learning}
\bibfield{author}{\bibinfo{person}{Zhenyu Song}, \bibinfo{person}{Daniel~S
  Berger}, \bibinfo{person}{Kai Li}, \bibinfo{person}{Anees Shaikh},
  \bibinfo{person}{Wyatt Lloyd}, \bibinfo{person}{Soudeh Ghorbani},
  \bibinfo{person}{Changhoon Kim}, \bibinfo{person}{Aditya Akella},
  \bibinfo{person}{Arvind Krishnamurthy}, \bibinfo{person}{Emmett Witchel},
  {et~al\mbox{.}}} \bibinfo{year}{2020}\natexlab{}.
\newblock \showarticletitle{Learning relaxed belady for content distribution
  network caching}. In \bibinfo{booktitle}{\emph{17th USENIX Symposium on
  Networked Systems Design and Implementation (NSDI 20)}}.
  \bibinfo{pages}{529--544}.
\newblock


\bibitem[Stuedi et~al\mbox{.}(2017)]%
        {stuedi2017crail}
\bibfield{author}{\bibinfo{person}{Patrick Stuedi}, \bibinfo{person}{Animesh
  Trivedi}, \bibinfo{person}{Jonas Pfefferle}, \bibinfo{person}{Radu Stoica},
  \bibinfo{person}{Bernard Metzler}, \bibinfo{person}{Nikolas Ioannou}, {and}
  \bibinfo{person}{Ioannis Koltsidas}.} \bibinfo{year}{2017}\natexlab{}.
\newblock \showarticletitle{Crail: A High-Performance I/O Architecture for
  Distributed Data Processing.}
\newblock \bibinfo{journal}{\emph{IEEE Data Eng. Bull.}} \bibinfo{volume}{40},
  \bibinfo{number}{1} (\bibinfo{year}{2017}), \bibinfo{pages}{38--49}.
\newblock


\bibitem[Vishwanath and Nagappan(2010)]%
        {vishwanath2010characterizing}
\bibfield{author}{\bibinfo{person}{Kashi~Venkatesh Vishwanath} {and}
  \bibinfo{person}{Nachiappan Nagappan}.} \bibinfo{year}{2010}\natexlab{}.
\newblock \showarticletitle{Characterizing cloud computing hardware
  reliability}. In \bibinfo{booktitle}{\emph{Proceedings of the 1st ACM
  symposium on Cloud computing}}. \bibinfo{pages}{193--204}.
\newblock


\bibitem[Wu and Buyya(2015)]%
        {wu2015cloud-diskspin}
\bibfield{author}{\bibinfo{person}{Caesar Wu} {and} \bibinfo{person}{Rajkumar
  Buyya}.} \bibinfo{year}{2015}\natexlab{}.
\newblock \bibinfo{booktitle}{\emph{Cloud Data Centers and Cost Modeling: A
  complete guide to planning, designing and building a cloud data center}}.
\newblock \bibinfo{publisher}{Morgan Kaufmann}.
\newblock


\bibitem[Yang et~al\mbox{.}(2019)]%
        {yang2019apachenemo}
\bibfield{author}{\bibinfo{person}{Youngseok Yang}, \bibinfo{person}{Jeongyoon
  Eo}, \bibinfo{person}{Geon-Woo Kim}, \bibinfo{person}{Joo~Yeon Kim},
  \bibinfo{person}{Sanha Lee}, \bibinfo{person}{Jangho Seo},
  \bibinfo{person}{Won~Wook Song}, {and} \bibinfo{person}{Byung-Gon Chun}.}
  \bibinfo{year}{2019}\natexlab{}.
\newblock \showarticletitle{Apache nemo: A framework for building distributed
  dataflow optimization policies}. In \bibinfo{booktitle}{\emph{2019 USENIX
  Annual Technical Conference (USENIX ATC 19)}}. \bibinfo{pages}{177--190}.
\newblock


\bibitem[Zaharia et~al\mbox{.}(2012)]%
        {zaharia2012resilient}
\bibfield{author}{\bibinfo{person}{Matei Zaharia}, \bibinfo{person}{Mosharaf
  Chowdhury}, \bibinfo{person}{Tathagata Das}, \bibinfo{person}{Ankur Dave},
  \bibinfo{person}{Justin Ma}, \bibinfo{person}{Murphy McCauly},
  \bibinfo{person}{Michael~J Franklin}, \bibinfo{person}{Scott Shenker}, {and}
  \bibinfo{person}{Ion Stoica}.} \bibinfo{year}{2012}\natexlab{}.
\newblock \showarticletitle{Resilient distributed datasets: A
  $\{$Fault-Tolerant$\}$ abstraction for $\{$In-Memory$\}$ cluster computing}.
  In \bibinfo{booktitle}{\emph{9th USENIX Symposium on Networked Systems Design
  and Implementation (NSDI 12)}}. \bibinfo{pages}{15--28}.
\newblock


\bibitem[Zaharia et~al\mbox{.}(2010)]%
        {zaharia2010spark}
\bibfield{author}{\bibinfo{person}{M Zaharia}, \bibinfo{person}{M Chowdhury},
  \bibinfo{person}{MJ Franklin}, \bibinfo{person}{S Shenker}, {and}
  \bibinfo{person}{I Stoica}.} \bibinfo{year}{2010}\natexlab{}.
\newblock \showarticletitle{Spark: Cluster computing with working sets. In:
  Proceedings of the 2Nd USENIX Conference on Hot Topics in Cloud Computing.
  HotCloud’10}.
\newblock  (\bibinfo{year}{2010}).
\newblock


\bibitem[Zhang et~al\mbox{.}(2018)]%
        {zhang2018riffle}
\bibfield{author}{\bibinfo{person}{Haoyu Zhang}, \bibinfo{person}{Brian Cho},
  \bibinfo{person}{Ergin Seyfe}, \bibinfo{person}{Avery Ching}, {and}
  \bibinfo{person}{Michael~J Freedman}.} \bibinfo{year}{2018}\natexlab{}.
\newblock \showarticletitle{Riffle: optimized shuffle service for large-scale
  data analytics}. In \bibinfo{booktitle}{\emph{Proceedings of the Thirteenth
  EuroSys Conference}}. \bibinfo{pages}{1--15}.
\newblock


\bibitem[Zhou and Maas(2021)]%
        {MLSYS2021_82161242}
\bibfield{author}{\bibinfo{person}{Giulio Zhou} {and} \bibinfo{person}{Martin
  Maas}.} \bibinfo{year}{2021}\natexlab{}.
\newblock \showarticletitle{Learning on Distributed Traces for Data Center
  Storage Systems}. In \bibinfo{booktitle}{\emph{Proceedings of Machine
  Learning and Systems}}, \bibfield{editor}{\bibinfo{person}{A.~Smola},
  \bibinfo{person}{A.~Dimakis}, {and} \bibinfo{person}{I.~Stoica}} (Eds.),
  Vol.~\bibinfo{volume}{3}. \bibinfo{pages}{350--364}.
\newblock
\urldef\tempurl%
\url{https://proceedings.mlsys.org/paper/2021/file/82161242827b703e6acf9c726942a1e4-Paper.pdf}
\showURL{%
\tempurl}


\end{thebibliography}

\end{document}